\documentclass{article}
\usepackage{spconf,epsfig}
\usepackage{times}
\usepackage{amsmath,amsfonts,bm,bbm,amssymb,framed}
\usepackage{graphicx,psfrag,color}
\usepackage[english]{babel}
\usepackage[numbers,sort]{natbib}
\bibpunct{[}{]}{,}{n}{}{;}
\usepackage{multirow}
\usepackage{multicol}
\usepackage{subfig}
  
\usepackage{tikz}
\usetikzlibrary{fit}
\usetikzlibrary{calc}
\usetikzlibrary{automata}

\graphicspath{{figures/}}

\DeclareMathAlphabet\PazoBB{U}{fplmbb}{m}{n}

\newcommand \Kcal {\mathcal{K}}

\newcommand{\vect}[1]{\mathbf{#1}}

\newcommand{\un}{\PazoBB{1}}
\newcommand{\model}{\mathcal{M}}
\newcommand{\kmax}{k_{\text{max}}}

\newcommand{\distrGauss}{\mathcal{N}}

\newcommand \ak  {\vect{a}_k}

\newcommand \ok  {\bm{\omega}_k}
\newcommand \thk {\bm{\theta}_k}
\newcommand \D   {\vect{D}}

\newcommand \y   {\vect{y}}

\newcommand{\dB}{\, \mathrm{dB}}
\newcommand{\SNR}{\mathrm{SNR}}

\newcommand{\E}{\bm{\eta}} 

\newcommand{\m}{\bm{\mu}} 
\newcommand{\s}{\bm{\Sigma}} 
\newcommand{\I}{\bm{\xi}} 

\newcommand{\XX}{\mathbb{X}}
\newcommand{\x}{\vect{x}}
\newcommand{\z}{\vect{z}}
\newcommand{\post}{f}
\newcommand \Nset {\mathbb{N}}
\newcommand \Rset {\mathbb{R}}

\newcommand \xxM [1]{\x^{(#1)}}
\newcommand \zzM [1]{\z^{(#1)}}

\newcommand{\mysubsec}[1]{\vspace{-.7mm}\subsection{#1}}

\DeclareMathOperator{\argmax}{argmax}

\title{SUMMARIZING POSTERIOR DISTRIBUTIONS IN SIGNAL DECOMPOSITION
  PROBLEMS WHEN THE NUMBER OF COMPONENTS IS UNKNOWN}
\name{Alireza Roodaki, Julien Bect, and Gilles Fleury}
\address{ E3S---SUPELEC Systems Sciences\\
  Dept. of Signal Processing and Electronic Systems, SUPELEC, Gif-sur-Yvette, France.\\
  Email: \{alireza.roodaki, julien.bect, gilles.fleury\}@supelec.fr}
\begin{document}
\ninept
\maketitle


\begin{abstract} 

  This paper addresses the problem of summarizing the posterior
  distributions that typically arise, in a Bayesian framework, when
  dealing with signal decomposition problems with unknown number of
  components. Such posterior distributions are defined over union of
  subspaces of differing dimensionality and can be sampled from using
  modern Monte Carlo techniques, for instance the increasingly popular
  RJ-MCMC method. No generic approach is available, however, to
  summarize the resulting variable-dimensional samples and extract
  from them component-specific parameters.
  
  We propose a novel approach to this problem, which consists in
  approximating the complex posterior of interest by a
  ``simple''---but still variable-dimensional---parametric
  distribution. The distance between the two distributions is measured
  using the Kullback-Leibler divergence, and a Stochastic EM-type
  algorithm, driven by the RJ-MCMC sampler, is proposed to estimate
  the parameters. The proposed algorithm is illustrated on the
  fundamental signal processing example of joint detection and
  estimation of sinusoids in white Gaussian noise.

\end{abstract}



\section{Introduction}\label{sec:intro}

Nowadays, owing to the advent of Markov Chain Monte Carlo (MCMC)
sampling methods~\cite{robert:2004:monte}, Bayesian data analysis is
considered as a conventional approach in machine learning, signal and
image processing, and data mining problems---to name but a
few. Nevertheless, in many applications, practical challenges remain
in the process of extracting, from the generated samples, quantities
of interest to summarize the posterior distribution.

Summarization consists, loosely speaking, in providing a few simple
yet interpretable parameters and/or graphics to the end-user of a
statistical method. For instance, in the case of a scalar parameter
with a unimodal posterior distribution, measures of location and
dispersion (e.g., the empirical mean and the standard deviation, or
the median and the interquartile range) are typically provided in
addition to a graphical summary of the distribution (e.g., a histogram
or a kernel density estimate). In the case of multimodal distributions
summarization becomes more difficult but can be carried out using, for
instance, the approximation of the posterior by a Gaussian Mixture
Model (GMMs) \cite{west:1993:approx}.

This paper addresses the problem of summarizing posterior
distributions in the case of trans-dimensional problems, i.e. ``the
problems in which the number of things that we don't know is one of
the things that we don't know''~\cite{green:1995:reversible}. The
problem of signal decomposition when the number of components is
unknown is an important example of such problems. Let
$\y\,=\,(y_1,\,y_2,\,\ldots,\,y_N)^t$ be a vector of $N$ observations,
where the superscript $t$ stands for vector transposition. In signal
decomposition problems, the model space is a finite or countable set
of models, $\model = \{ \model_k,\, k \in \Kcal \}$, where $\Kcal
\subset \Nset$ is an index set. It is assumed here that,
under~$\model_k$, there are $k$ components with component-specific
parameters $\bm{\theta}_{k,j} \in \bm{\Theta} \subset
\mathbb{R}^d$. We denote by~$\thk = \left( \bm{\theta}_{k,1},
  \ldots, \bm{\theta}_{k,k} \right) \in \bm{\Theta}^k$ the vector of
component-specific parameters. In a Bayesian framework,
a joint posterior distribution is obtained through Bayes' formula for
the model index~$k$ and the vector of component-specific parameters,
after assigning prior distributions on them :
\begin{align*}
  \post\left(k,\,\thk \right) 
  \;\propto\;
  p\left(\y\,|\,k,\,\thk\right)p\left(\thk\,|\,k\right)p\left(k\right),
\end{align*}
where $\propto$ indicates proportionality. This joint posterior
distribution, defined over a union of subspaces of differing
dimensionality, completely describes the information (and the
associated uncertainty) provided by the data~$\y$ about the candidate
models and the vector of unknown parameters.

\mysubsec{Illustrative example: sinusoid detection}\label{sec:Example}
 
In this example, it is assumed that under $\model_k$, $\y$ can be
written as a linear combination of~$k$ sinusoids observed in white
Gaussian noise. The unknown component-specific parameters are
$\thk=\{\ak, \ok, \bm{\phi}_k\}$, where $\ak$, $\ok$
and~$\bm{\phi}_k$ are the vectors of amplitudes, radial frequencies
and phases, respectively.  We use the hierarchical model, prior
distributions, and Reversible Jump MCMC (RJ-MCMC)
sampler~\cite{green:1995:reversible} proposed
in~\cite{andrieu:1999:jbm} for this problem; the interested reader is
thus referred to~\cite{andrieu:1999:jbm, green:1995:reversible} for
more details.

\begin{figure}  
  \begin{center} 
%
%
\begin{psfrags}%
\psfragscanon%
\newcommand{\Xtick}[2]{\psfrag{#1}[t][t]{\footnotesize #2}}
\newcommand{\Ytick}[2]{\psfrag{#1}[r][r]{\footnotesize #2}}
\psfrag{s06}[t][t][0.75]{\setlength{\fboxsep}{2pt}\fcolorbox[rgb]{0,0,0}{1,1,1}{\color[rgb]{0,0,0}\setlength{\tabcolsep}{0pt}\begin{tabular}{l}$59.5\%$
 \end{tabular}}}%
\psfrag{s07}[t][t][0.75]{\setlength{\fboxsep}{2pt}\fcolorbox[rgb]{0,0,0}{1,1,1}{\color[rgb]{0,0,0}\setlength{\tabcolsep}{0pt}\begin{tabular}{l}$30.8\%$ 
\end{tabular}}}%
\psfrag{s08}[t][t][0.75]{\setlength{\fboxsep}{2pt}\fcolorbox[rgb]{0,0,0}{1,1,1}{\color[rgb]{0,0,0}\setlength{\tabcolsep}{0pt}\begin{tabular}{l}$7.8\%$
 \end{tabular}}}%
\psfrag{s09}[b][b][0.75]{\color[rgb]{0,0,0}\setlength{\tabcolsep}{0pt}\begin{tabular}{c}$k$\end{tabular}}%
\psfrag{s10}[l][l]{\setlength{\fboxsep}{2pt}\fcolorbox[rgb]{0,0,0}{0.8,0.8,0.8}{\color[rgb]{0,0,0}\setlength{\tabcolsep}{0pt}\begin{tabular}{l}$59.5$
 \end{tabular}}}%
\psfrag{s11}[l][l]{\setlength{\fboxsep}{2pt}\fcolorbox[rgb]{0,0,0}{0.8,0.8,0.8}{\color[rgb]{0,0,0}\setlength{\tabcolsep}{0pt}\begin{tabular}{l}$30.8$
 \end{tabular}}}%
\psfrag{s12}[l][l]{\setlength{\fboxsep}{2pt}\fcolorbox[rgb]{0,0,0}{0.8,0.8,0.8}{\color[rgb]{0,0,0}\setlength{\tabcolsep}{0pt}\begin{tabular}{l}$7.8$
 \end{tabular}}}%
\psfrag{s14}[t][t][0.75]{\color[rgb]{0,0,0}\setlength{\tabcolsep}{0pt}\begin{tabular}{c}$\ok$\end{tabular}}%
%
\Xtick{x01}{$0.5$}%
\Xtick{x02}{$0.75$}%
\Xtick{x03}{$1$}%
%
\Ytick{v01}{$2$}%
\Ytick{v02}{$3$}%
\Ytick{v03}{$4$}%
%
\includegraphics[width=78mm]{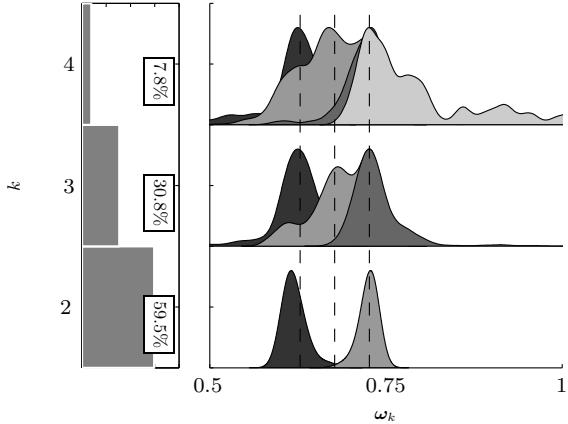}%
\end{psfrags}%
%
 
    \vspace{-1mm}
    \caption{Posteriors of $k$ (left) and sorted radial frequencies,
      $\ok$, given $k$ (right). The true number of components is
      three. The vertical dashed lines in the right figure locate the
      true radial frequencies.}
    \label{fig:visu}
  \end{center}
\end{figure}

Figure~\ref{fig:visu} represents the posterior distributions of both
the number of components~$k$ and the sorted\footnote{Owing to the
  invariance of both the likelihood and the prior under permutation of
  the components, component-specific marginal posteriors are all
  equal: this is the ``label-switching''
  phenomenon~\cite{richardson:1997:bayesian, stephens:2000:label,
    sperrin:2010:label}. Identifiability constraints (such as sorting)
  are the simplest way of dealing with this issue.}  radial
frequencies~$\ok$ given~$k$ obtained using the RJ-MCMC sampler. Each
row is dedicated to one value of $k$, for $2\leq k\leq4$; observe
that, other models have negligible posterior probabilities. In the
experiment, the observed signal of length $N=64$ consists of three
sinusoids with amplitudes $\ak=(20, 6.32, 20)^t$
and radial frequencies $\ok=(0.63, 0.68, 0.73)^t$. The
$\SNR\triangleq\frac{\|\D_k .\ak\|^2}{N\sigma^2}$ is set to the
moderate value of~$7\dB$, where $\D_k$ is the design matrix and
$\sigma^2$ is the noise variance.

Roughly speaking, two approaches co-exist in the literature for such
situations: Bayesian Model Selection (BMS) and Bayesian Model
Averaging (BMA). The BMS approach ranks models according to their
posterior probabilities $p(k|\y)$, selects one model, and then
summarizes the posterior under the (fixed-dimensional) selected model.
This is at the price of loosing valuable information provided by the
other (discarded) models. For instance, in the example of
Figure~\ref{fig:visu}, all information about the small---and therefore
harder to detect---middle component is lost by selecting the most
\emph{a posteriori} probable model~$\model_2$. The BMA approach
consists in reporting results that are averaged over all possible
models; it is, therefore, not appropriate for studying
component-specific parameters, the number of which changes in each
model\footnote{See, however, the intensity plot provided in
  Section~\ref{sec:result} (middle plot on Figure~\ref{fig:intens}) as
  an example of a BMA summary related to a component-specific
  parameter.}.

More information concerning these two approaches can be found
in~\cite{green:1995:reversible} and references therein. To the best of
our knowledge, no generic method is currently available, that would
allow to summarize the information that is so easily read on
Figure~\ref{fig:visu} for this very simple example: namely, that
\emph{there seem to be three sinusoidal components in the observed
  noisy signal, the middle one having a smaller probability of
  presence than the others}.

\mysubsec{Outline of the paper}\label{sec:outline}

In this paper, we propose a novel approach to summarize the posterior
distributions over variable-dimensional subspaces that typically arise
in signal decomposition problems with an unknown number of components.
It consists in approximating the complex posterior distribution with a
parametric model (of varying-dimensionality), by minimization of the
Kullback-Leibler (KL) divergence between the two distributions. A
Stochastic EM (SEM)-type algorithm \cite{celeux:1985:SEM}, driven by
the output of an RJ-MCMC sampler, is used to estimate the parameters
of the approximate model.

Our approach shares some similarities with the relabeling algorithms
proposed in~\cite{stephens:2000:label, sperrin:2010:label} to solve
the ``label switching'' problem, and also with the EM algorithm used
in~\cite{bai:2011:divcon} in the context of adaptive MCMC algorithms
(both in a \emph{fixed}-dimensional setting). The main contribution of
this paper is the introduction of an original variable-dimensional
parametric model, which allows to tackle directly the difficult
problem of approximating a distribution defined over a union of
subspaces of differing dimensionality---and thus provides a first
solution to the ``trans-dimensional label-switching'' problem, so to
speak.

The paper is organized as follows. Section~\ref{sec:TAP} introduces
the proposed model and stochastic algorithm. Section~\ref{sec:result}
illustrates the approach using the sinusoid detection example already
discussed in the introduction. Finally, Section~\ref{sec:conclusion}
concludes the paper and gives directions for future work.

\section{Proposed algorithm} \label{sec:TAP}

Let~$F$ denote the target posterior distribution, defined on the
variable-dimensional space~$\XX = \bigcup_{k=0}^{\kmax}\, \{k\} \times
\bm{\Theta}^k$. We assume that~$F$ admits a probability density
function (pdf) $\post$, with respect to the $kd$-dimensional Lebesgue
measure on each~$\{k\} \times \bm{\Theta}^k$, $k \in \Kcal$. To keep
things simple, we also assume that $\bm{\Theta} = \Rset^d$.

Our objective is to approximate the exact posterior density~$f$ using
a ``simple'' parametric model~$q_{\E}$, where $\E$ is the vector of
parameters defining the model. The pdf~$q_{\E}$ will \emph{also} be
defined on the variable-dimensional space~$\XX$ (i.e., it is not a
fixed-dimensional approximation as in the BMS approach). We assume
that a Monte Carlo sampling method is available, e.g. a RJ-MCMC
sampler~\cite{green:1995:reversible}, to generate~$M$ samples
from~$\post$, which we denote by~$\xxM{i} = \bigl( k^{(i)},
\bm{\theta}^{(i)}_{k^{(i)}} \bigr)$, for $i \,=\, 1, \ldots, M$.

\mysubsec{Variable-dimensional parametric model}\label{sec:model}

Let us describe the proposed parametric model from a generative point
of view.  As in a traditional GMM, we assume that there is a certain
number~$L$ of ``Gaussian components'' in the (approximate) posterior,
each generating a $d$-variate Gaussian vector with mean~$\m_l$ and
covariance matrix~$\s_l$, $1 \le l \le L$.

An $\XX$-valued random variable~$\x = ( k, \bm{\theta}_k )$, with $0
\le k \le L$, is generated as follows. First, each of the~$L$
components can be either present or absent according to a binary
indicator variable~$\xi_l \in \{ 0, 1 \}$. These Bernoulli variables
are assumed to be independent, and we denote by $\pi_l \in (0;1]$ the
``probability of presence'' of the~$l^{\text{th}}$ Gaussian
component. Second, given the indicator variables, $k = \sum_{l=1}^L
\xi_l$ Gaussian vectors are generated by the Gaussian components that
are present ($\xi_l = 1$) and randomly arranged in a vector
$\bm{\theta}_k = \left( \bm{\theta}_{k,1}, \ldots, \bm{\theta}_{k,k}
\right)$.

We denote by $q_{\E}$ the pdf of the random variable~$\x$ that is thus
generated, with $\E_l = \left( \m_l, \s_l, \pi_l \right)$ the vector
of parameters of the $l^{\text{th}}$ Gaussian component, $1 \le l \le
L$, and $\E = \left( \E_1, \ldots, \E_L \right)$.

\medbreak

\noindent\textbf{Remark.} In contrast with GMMs, where only one component is
present at a time (i.e., $k = 1$ in our notations), there is no
constraint here on the sum of the probabilities of presence.

\mysubsec{Estimating the model parameters}\label{sec:estim}

One way to fit the parametric distribution~$q_{\E}(\x)$ to the
posterior~$\post(\x)$ is to minimize the KL divergence of $\post$ from
$q_{\E}$, denoted by $D_{KL}(\post(\x) \| q_{\E}(\x))$. Thus, we
define the criterion to be minimized as
\begin{equation*} 
  \mathcal{J}(\E) \; \triangleq \; D_{KL}\left(\post(\x) \| q_{\E}(\x)\right)
  \;=\;\int_{\XX} \post\left(\x\right) \,\log\frac{\post(\x)}{q_{\E}(\x)} \;\text{d}\x.
\end{equation*}
Using samples generated by the RJ-MCMC sampler, this criterion can be
approximated as
\begin{equation*}
  \mathcal{J}(\E) \; \simeq \; \hat{\mathcal{J}}(\E)
  \;=\; -\frac{1}{M}\sum_{i=1}^M \,\log\left(q_{\E}(\xxM{i})\right) + C.
\end{equation*}
where $C$ is a constant that does not depend on~$\E$.  One should note
that minimizing $\hat{\mathcal{J}}(\E)$ amounts to estimating $\E$
such that
\begin{equation}
  \label{equ:MLE}
  \hat{\E} \;=\; \argmax_{\E}
  \sum_{i=1}^M \log\left(q_{\E}(\xxM{i})\right).
\end{equation}

Now, we assume that each element of the~$i^{\text{th}}$ observed
sample~$\xxM{i}_j$, for~$j\,=\,1,\,\ldots,\,k^i$, has arisen from one
of the~$L$ Gaussian components contained in~$q_{\E}$. At this point,
it is natural to introduce allocation vectors $\zzM{i}$ corresponding
to the~$i^{\text{th}}$ observed sample~$\xxM{i}$, for $i \,=\,
1,\ldots,M$, as latent variables. The element $\zzM{i}_j=l$ indicates
that $\xxM{i}_j$ is allocated to the~$l^{\text{th}}$ Gaussian
component.

Hence, given the allocation vector~$\zzM{i}$ and the parameters of the
model~$\E$, the conditional distribution of the observed samples,
i.e., the model's likelihood, is
\begin{equation*}\label{eq: obs}
  p(\xxM{i}\,|\,\zzM{i},\,\E) 
  \,=\,
  \prod_{j=1}^{k^{(i)}} \distrGauss(\xxM{i}_j\,|\,\m_{\zzM{i}_j},\,\s_{\zzM{i}_j}).
\end{equation*}

It turns out that the EM-type algorithms, which have been used in
similar works~\cite{stephens:2000:label, sperrin:2010:label,
  bai:2011:divcon}, are not appropriate for solving this problem, as
computing the expectation in the E-step is intricate. More explicitly,
in our problem the computational burden of the summation in the E-step
over the set of all possible allocation vectors~$\z$ increases very
rapidly with~$k$. In fact, even for moderate values of~$k$, say, $k =
10$, the summation is far too expensive to compute as it involves $k!
\approx 3.6\, 10^6$ terms. In this paper, we propose to use
SEM~\cite{celeux:1985:SEM}, a variation of the EM algorithm in which
the E-step is substituted with stochastic simulation of the latent
variables from their conditional posterior distributions given the
previous estimates of the unknown parameters. In other words, for
$i\,=\,1,\,\ldots,\,M$, the allocation vectors $\zzM{i}$ are drawn
from $p(\bm{\cdot}\,|\,\xxM{i},\,\hat{\E}^{(r)})$. This step is called
the Stochastic (S)-step. Then, these random samples are used to
construct the so-called pseudo-completed likelihood which reads
\begin{align}\label{eq:comp-like}
  p \left(\xxM{i}, \zzM{i} \,|\,\E \right) & \;=\;
  \prod_{j=1}^{k^{(i)}} \distrGauss \left( \xxM{i}_j%
    \,|\,
    \m_{\zzM{i}_j},\,\s_{\zzM{i}_j} \right) \nonumber\\
  &\times %
  \frac{ %
    \un_{ \mathcal{Z} } (\zzM{i})%
  } %
  {k^{(i)}!}%
  \,\prod_{l=1}^L%
  \pi_l^{\I^{(i)}_l} \left( 1 \,-\, \pi_l \right)^{(1-\I^{(i)}_l)},
\end{align} 
where $\mathcal{Z}$ is the set of all allocation vectors and
$\I^{(i)}_l = 1$ if and only if there is a $j \in \{ 1, \ldots,
k^{(i)} \}$ such that $\zzM{i}_j = l$. The proposed SEM-type algorithm
for our problem is described in Figure~\ref{fig: SEM}.
 
\begin{figure}[t]
  \begin{framed} 
    At the $r^{\text{th}}$ iteration,
    \begin{description}

    \item[S-step] draw allocation vectors $\zzM{i,r} \sim p
      \left(\, \bm{\cdot} \,|\, \xxM{i} ,\, \hat{\E}^{(r-1)} \right)$, for
      $i\,=\,1,\,\ldots,\,M$.
 
    \item[M-step] estimate~$\hat{\E}^{(r)}$ such that \vspace{-10pt}
      \begin{equation*}
        \hat{\E}^{(r)}\;=\;\argmax_{\E}\sum_{i=1}^M \log \, p \left(\xxM{i}, \zzM{i,r} \,|\,\E \right).
      \end{equation*}
    \end{description}
    \vspace{-12pt}
  \end{framed}
  \vspace{-1em}
  \caption{SEM algorithm.}
  \label{fig: SEM}
\end{figure}

Direct sampling from $p(\, \bm{\cdot} \,|\, \xxM{i} ,\, \hat{\E}^{(r)}
)$, as required by the S-step, is unfortunately not feasible. Instead,
since
\begin{equation*}
  p (\zzM{i} \,|\, \xxM{i}, \, \hat{\E}^{(r)} ) 
  \;\propto\;
  p (\xxM{i},\zzM{i} \,|\, \hat{\E}^{(r)} )  
\end{equation*}
can be computed up to a normalizing constant, we devised an
Independent Metropolis-Hasting (I-MH) algorithm to construct a Markov
chain with $p (\zzM{i} \,|\, \xxM{i},\,\hat{\E}^{(r)} )$ as its
stationary distribution.

\mysubsec{Robustified algorithm}\label{sec:robust}

Preliminary experiments with the model and method described in the
previous sections proved to be disappointing. To understand why, it
must be remembered that the pdf~$q_{\E}$ we are looking for is only an
\emph{approximation} (hopefully a good one) of the true
posterior~$f$. For instance, for high values of~$k$, the posterior
typically involves a diffuse part which can not properly represented
by the parametric model (this can be seen quite clearly for $k=4$ on
Figure~\ref{fig:visu}). Therefore, for any~$\E$, some samples
generated by the RJ-MCMC sampler are \emph{outliers} with respect
to~$q_{\E}$ (i.e., the true posterior can be considered as a
\emph{contaminated} version of~$q_{\E}$) which causes problems when
using a maximum likelihood-type estimate such as~\eqref{equ:MLE}.

These robustness issues were solved, in this paper, using two
modifications of the algorithm (only in the one-dimensional case up to
now). First, robust estimates \cite{huber:2009:robust} of the means
and variances of a Gaussian distribution, based on the median and the
interquartile range, are used instead of the empirical means and
variances in the M-step. Second, a Poisson process component (with
uniform intensity) is added to the model, in order to account for the
diffuse part of the posterior and allow for a number~$L$ of Gaussian
components which is smaller than the maximum observed~$k^{(i)}$.

\medbreak\noindent\textbf{Remark.} Similar robustness concerns are
widespread in the clustering literature; see, e.g.,
\cite{dave1997robust} and the references therein.

\section{Results} \label{sec:result}

In this section, we will investigate the capability of the proposed
algorithm for summarizing variable-dimensional posterior
distributions. We emphasize again that the output of the
trans-dimensional Monte Carlo sampler, e.g. RJ-MCMC in this paper, is
considered as the observed data for our algorithm. Regarding the fact
that in this paper we provide results for the sinusoids' radial
frequencies, the proposed parametric model consists of univariate
Gaussian components. In other words, the space of component-specific
parameters $\bm{\Theta} = (0; \pi) \subset \Rset$. But we believe that
our algorithm is not limited to the problems with one-dimensional
component-specific parameters. Therefore, in this section, it is
assumed that each Gaussian component has a mean~$\mu$, a
variance~$s^2$, and a probability of presence~$\pi$ to be estimated.

Before launching the algorithm, first, we need to initialize the
parametric model.  It is natural to deduce the number~$L$ of Gaussian
components from the posterior distribution of~$k$. Here, we set it to
the $90^{th}$ percentile to keep all the probable models in the
play. To initialize the Gaussian components' parameters, i.e. $\mu$
and $s^2$, we used the robust estimates of the posterior of the sorted
radial frequencies given~$k=L$.

\begin{figure}
 \centering 
%
%
\begin{psfrags}%
\psfragscanon%
\newcommand{\Xtick}[2]{\psfrag{#1}[t][t]{\footnotesize #2}}
\newcommand{\Ytick}[2]{\psfrag{#1}[r][r]{\footnotesize #2}}
\psfrag{s11}[b][b][0.75]{\color[rgb]{0,0,0}\setlength{\tabcolsep}{0pt}\begin{tabular}{c}$\mu$\end{tabular}}%
\psfrag{s12}[b][b][0.75]{\color[rgb]{0,0,0}\setlength{\tabcolsep}{0pt}\begin{tabular}{c}$s^2$\end{tabular}}%
\psfrag{s13}[b][b][0.75]{\color[rgb]{0,0,0}\setlength{\tabcolsep}{0pt}\begin{tabular}{c}$\pi$\end{tabular}}%
\psfrag{s14}[t][t][0.75]{\color[rgb]{0,0,0}\setlength{\tabcolsep}{0pt}\begin{tabular}{c}SEM iteration\end{tabular}}%
\psfrag{s15}[b][b][0.75]{\color[rgb]{0,0,0}\setlength{\tabcolsep}{0pt}\begin{tabular}{c}$J$\end{tabular}}%
\psfrag{s16}[t][t][0.75]{\color[rgb]{0,0,0}\setlength{\tabcolsep}{0pt}\begin{tabular}{c}SEM iteration\end{tabular}}%
%
\Xtick{x01}{$10$}%
\Xtick{x02}{$20$}%
\Xtick{x03}{$30$}%
\Xtick{x04}{$40$}%
\Xtick{x05}{$50$}%
\Xtick{x06}{$10$}%
\Xtick{x07}{$20$}%
\Xtick{x08}{$30$}%
\Xtick{x09}{$40$}%
\Xtick{x10}{$50$}%
\Xtick{x11}{$0$}%
\Xtick{x12}{$5$}%
\Xtick{x13}{$10$}%
\Xtick{x14}{$15$}%
\Xtick{x15}{$20$}%
\Xtick{x16}{$25$}%
\Xtick{x17}{$30$}%
\Xtick{x18}{$35$}%
\Xtick{x19}{$40$}%
\Xtick{x20}{$45$}%
\Xtick{x21}{$50$}%
\Xtick{x22}{$0$}%
\Xtick{x23}{$5$}%
\Xtick{x24}{$10$}%
\Xtick{x25}{$15$}%
\Xtick{x26}{$20$}%
\Xtick{x27}{$25$}%
\Xtick{x28}{$30$}%
\Xtick{x29}{$35$}%
\Xtick{x30}{$40$}%
\Xtick{x31}{$45$}%
\Xtick{x32}{$50$}%
%
\Ytick{v01}{$1$}%
\Ytick{v02}{$2$}%
\Ytick{v03}{$3$}%
\Ytick{v04}{$0.25$}%
\Ytick{v05}{$0.5$}%
\Ytick{v06}{$0.75$}%
\Ytick{v07}{$1$}%
\Ytick{v08}{$10^{-4}$}%
\Ytick{v09}{$10^{-3}$}%
\Ytick{v10}{$10^{-2}$}%
\Ytick{v11}{$0.6$}%
\Ytick{v12}{$0.66$}%
\Ytick{v13}{$0.72$}%
%
\includegraphics[width=7.5cm]{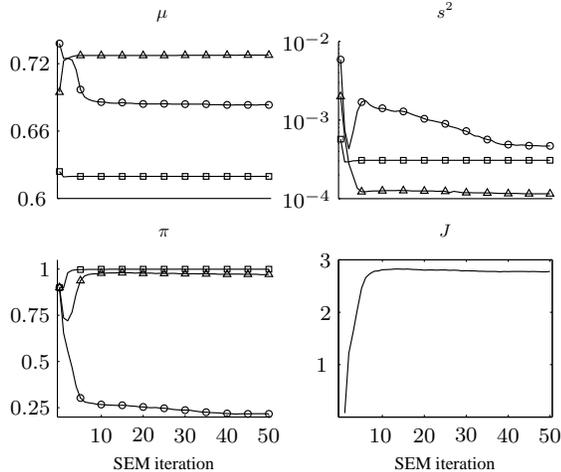}%
\end{psfrags}%
%

 \caption{Performance of the proposed summarizing algorithm on the
   sinusoid detection example. There are three Gaussian components in
   the model.}
  \label{fig:param_evol}
\end{figure}

We ran the ``robustified'' stochastic algorithm introduced in
Section~\ref{sec:TAP} on the specific example shown in
Figure~\ref{fig:visu}, for 50 iterations, with $L = 3$ Gaussian
components (the posterior
probability of $\{ k \le 3 \}$ is approximately 90.3\%).
 Figure~\ref{fig:param_evol} illustrates the evolution of model
parameters~$\E$ together with the criterion~$\mathcal{J}$. Two
substantial facts can be deduced from this figure; first, the
increasing behavior of the criterion~$\mathcal{J}$, which is almost
constant after the $10^{th}$ iteration. Second, the convergence of the
parameters of parametric model, esp. means $\mu$ and probabilities of
presence $\pi$, though using a naive initialization procedure. Indeed
after the $40^{th}$ iteration there is no significant move in the
parameter estimates. Table~\ref{table:param} presents the summaries
provided by the proposed method along with the ones obtained using the
BMS approach. Contrary to BMS, the method that we proposed has enabled
us to benefit from the information of all probable models to give
summaries about the middle harder to detect component. Turning to the
results of our approach, it can be seen that the estimated means are
compatible with the true radial frequencies. Furthermore, the
estimated probabilities of presence are consistent with uncertainty of
them in the variable-dimensional posterior shown in
Figure~\ref{fig:visu}. Note the small estimated standard deviations
which indicate our robustifying strategies have been useful.

\begin{figure}
  \centering 
%
%
\begin{psfrags}%
\psfragscanon%
\newcommand{\Xtick}[2]{\psfrag{#1}[t][t]{\footnotesize #2}}
\newcommand{\Ytick}[2]{\psfrag{#1}[r][r]{\footnotesize #2}}
\psfrag{s06}[b][b][0.75]{\color[rgb]{0,0,0}\setlength{\tabcolsep}{0pt}\begin{tabular}{c}pdf\end{tabular}}%
\psfrag{s10}[b][b][0.75]{\color[rgb]{0,0,0}\setlength{\tabcolsep}{0pt}\begin{tabular}{c}intensity\end{tabular}}%
\psfrag{s11}[b][b][0.75]{\color[rgb]{0,0,0}\setlength{\tabcolsep}{0pt}\begin{tabular}{c}intensity\end{tabular}}%
\psfrag{s12}[t][t][0.75]{\color[rgb]{0,0,0}\setlength{\tabcolsep}{0pt}\begin{tabular}{c}$\ok$\end{tabular}}%
%
\Xtick{x01}{$0$}%
\Xtick{x02}{$1$}%
\Xtick{x03}{$2$}%
\Xtick{x04}{$3$}%
\Xtick{x05}{$0$}%
\Xtick{x06}{$0.5$}%
\Xtick{x07}{$1$}%
\Xtick{x08}{$1.5$}%
\Xtick{x09}{$2$}%
\Xtick{x10}{$2.5$}%
\Xtick{x11}{$3$}%
\Xtick{x12}{$0$}%
\Xtick{x13}{$0.5$}%
\Xtick{x14}{$1$}%
\Xtick{x15}{$1.5$}%
\Xtick{x16}{$2$}%
\Xtick{x17}{$2.5$}%
\Xtick{x18}{$3$}%
%
\Ytick{v01}{$0$}%
\Ytick{v02}{$1$}%
\Ytick{v03}{$2$}%
\Ytick{v04}{$0$}%
\Ytick{v05}{$10$}%
\Ytick{v06}{$20$}%
\Ytick{v07}{$0$}%
\Ytick{v08}{$10$}%
\Ytick{v09}{$20$}%
\Ytick{v10}{$30$}%
%
\includegraphics[width=8cm]{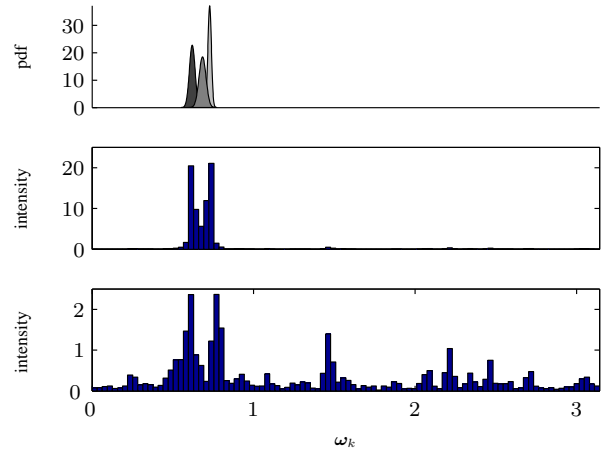}%
\end{psfrags}%
%
 
  \vspace{-2mm}
  \caption{The pdf of fitted Gaussian components (top), the histogram
    intensity of all radial frequencies samples (middle), and the
    histogram intensity of the allocated samples to the Poisson point
    process component (bottom).}
  \label{fig:intens}
\end{figure}

The pdf's of the estimated Gaussian components are shown in
Figure~\ref{fig:intens} (top).  Comparing with the posterior of sorted
radial frequencies shown in Figure~\ref{fig:visu}, it can be inferred
that the proposed algorithm has managed to remove the label-switching
phenomenon in a variable-dimensional problem. Furthermore, the
intensity plot of the allocated samples to the point process component
is depicted in Figure~\ref{fig:intens} (bottom). This presents the
outliers in the observed samples which cannot be be described by the
Gaussian components. Note that without the point process component
these outliers would be allocated to the Gaussian components which
can, consequently, yield in a significant deterioration of parameter
estimates.

\begin{table}
  \centering
  \begin{tabular}{|c|c|c|c|c|c|}
    \hline
    Comp&$\mu$&$s$&$\pi$&$\mu_{BMS}$&$s_{BMS}$\\ 
    \hline        
    $1$&$0.62$&$0.017$&$1$&$0.62$&$0.016$\\
    \hline
    $2$&$0.68$&$0.021$&$0.22$&---&---\\
    \hline
    $3$&$0.73$&$0.011$&$0.97$&$0.73$&$0.012$\\
    \hline
  \end{tabular}
  \caption{Summaries of the variable-dimensional posterior distribution shown in Figure~\ref{fig:visu}; The proposed approach vs. BMS. }
  \label{table:param}
\end{table} 

\section{Conclusion} \label{sec:conclusion} 

In this paper, we have proposed a novel algorithm to summarize
posterior distributions defined over union of subspaces of differing
dimensionality. For this purpose, a variable-dimensional parametric
model has been designed to approximate the posterior of interest. The
parameters of the approximate model have been estimated by means of a
SEM-type algorithm, using samples from the posterior~$\post$ generated
by an RJ-MCMC algorithm. Modifications of our initial SEM-type
algorithm have been proposed, in order to cope with the lack of
robustness of maximum likelihood-type estimates. The relevance of the
proposed algorithm, both for summarizing and for relabeling
variable-dimensional posterior distributions, has been illustrated on
the problem of detecting and estimating sinusoids in Gaussian white
noise. 

We believe that this algorithm can be used in the vast domain of
signal decomposition and mixture model analysis to enhance inference
in trans-dimensional problems.  For this purpose, generalizing the
proposed algorithm to the multivariate case and analyzing its
convergence properties is considered as future work. Another important
point would be to use a more reliable initialization procedure.

\bibliographystyle{IEEEbib}
\scriptsize\setlength{\bibsep}{0.5ex}
\bibliography{REF}

\end{document}